\documentclass[aps,pra,twocolumn,showpacs,superscriptaddress]{revtex4-1}

\usepackage[T1]{fontenc}
\usepackage[latin9]{inputenc}
\usepackage{amsmath}
\usepackage{amssymb}
\usepackage{ifthen}
\usepackage{times}
\usepackage{graphicx}
\usepackage{psfrag}

\def\be{\begin{equation}}
\def\ee{\end{equation}}
\def\bea{\begin{eqnarray}}
\def\eea{\end{eqnarray}}
\def\bi{\begin{itemize}}
\def\ei{\end{itemize}}

\usepackage{amssymb,amsmath}
\usepackage[usenames]{color}
\usepackage{graphicx}
\newcommand{\op}[1]{\widehat{#1}}
\newcommand{\dagop}[1]{\widehat{#1}^{\dagger}}
\newcommand{\bo}[1]{{\mathbf{#1}}}

\newcommand{\wt}[1]{{\widetilde{#1}}}

\newcommand{\nonu}{\nonumber}

\newlength{\templength}

\newcommand{\REM}[1]{\ifthenelse{0=1}{#1}{}}

\begin{document}

\setlength{\unitlength}{1cm}

\title{Mean field effects on the scattered atoms in condensate collisions}

\author{P. Deuar}
\affiliation{Institute of Physics, Polish Academy of Sciences, Al. Lotnik\'ow
32/46, 02-668 Warsaw, Poland}

\author{P. Zi\'n}
\affiliation{The Andrzej So{\l }tan Institute for Nuclear Studies,
Ho\.{z}a 69, PL-00-681 Warsaw, Poland}

\author{J. Chwede\'nczuk}
\affiliation{Institute of Theoretical Physics, Physics Department, University
of Warsaw, Ho\.{z}a 69, PL-00-681 Warsaw, Poland}

\author{M. Trippenbach}
\affiliation{Institute of Theoretical Physics, Physics Department, University
of Warsaw, Ho\.{z}a 69, PL-00-681 Warsaw, Poland}

\date{\today}
\begin{abstract}
  We consider the collision of two Bose Einstein condensates at supersonic velocities and focus on the halo of scattered atoms. 
  This halo 
  is the most important feature for experiments and is also an excellent
  testing ground
  for various theoretical approaches.
  In particular we find that the typical reduced Bogoliubov description, commonly used, is often not accurate in the region of parameters where  experiments are performed.
  Surprisingly, besides  the halo pair creation terms, one should take into account the evolving mean field of the remaining condensate and on-condensate pair creation.
  We present examples where the difference is clearly seen, and where the reduced description still holds.
\end{abstract}

\maketitle

\section{Introduction}
\label{INTRO}

When two Bose-Einstein condensates (BECs) collide at sufficiently high velocity, pairs of atoms are scattered out of the condensates.
After many scattering events, a distinct halo of atom pairs is formed in momentum space. This was observed in many experiments
\cite{Kozuma99,Deng99,Chikkatur00,Maddaloni00,Band01,Steinhauer02,Vogels02,Katz02,Vogels03,Katz04,Buggle04,Katz05,Perrin07,Perrin08,Dall09,Krachmalnicoff10,Jaskula10}
and analyzed in numerous theoretical works
 \cite{Band00,Trippenbach00,Yurovsky02,Bach02,Chwedenczuk04,Norrie05,Zin05,Katz05,Chwedenczuk06,Zin06,Norrie06,Deuar07,Drummond07,
Chwedenczuk08,Perrin08,Molmer08,Ogren09,Deuar09,Wang10,Krachmalnicoff10}.
The formation of a halo starts spontaneously and is analogous to the generation of photon pairs in parametric down conversion.
Such photon pairs were used to observe Bell inequality violation  \cite{Kwiat},  and can be applied for quantum cryptography  \cite{Gisin}
or quantum teleportation  \cite{Bou}. In analogy, atoms formed in the collision of two BECs have a potential application for precision measurements
 \cite{Bachor04}, interferometry
 \cite{Gross10,Bouyer97,Dunningham02,Campos03,Jaskula10}, or tests of quantum
mechanics  \cite{Reid09}.

The simplest model that captures the formation of a halo in the condensate collision  is  the ``reduced Bogoliubov'' model (RBM). In this formulation, the two condensates counter-propagate at a constant relative velocity
and without change of shape. The wave-packets enter in the Bogoliubov equation for the field of scattered atoms
as a classical source  \cite{Bach02,Yurovsky02,Zin05,Chwedenczuk06,Zin06,Chwedenczuk08,Trippenbach00,Band00,Band01,Chwedenczuk04,Ogren09}. The RBM can be used to calculate various observables, such as the density of scattered atoms or their second-order correlation functions, which can be directly compared with the experimental data.
The predictions of the RBM are often in good agreement with experiment \cite{Chwedenczuk08}, but there are cases where more complete models
have shown significant departures from the RBM  \cite{Deuar07,Ogren09,Krachmalnicoff10}.

In this paper we compare results of the RBM and the complete Bogoliubov equation for a range of parameters. In Section \ref{METH} 
we introduce both models and the numerical method of solving the
equations of motion. Two additional simplified formulations, also introduced in Section \ref{METH}, are used to investigate the properties of the halo.
The dependence of the halo shape on the model is shown in Section~\ref{EG1} for two characteristic cases, whereas Sec.~\ref{EG} explains in detail the dependence of the halo position
on the degree of simplifications made. We also indicate the range of interaction strength for which the RBM is quantitatively accurate.

\section{Bogoliubov approach to the scattering of atoms in a BEC collision}
\label{METH}

\subsection{Scattering in condensate collisions}

Initially the gas is trapped by a harmonic potential. In order to start the (half-) collision,
a superposition of two counter-propagating, mutually coherent, atomic clouds of equal density is prepared by shining a Bragg pulse.
The trap is simultaneously turned off. The two fractions move apart with
relative speed $2v_{\rm rec}$ along the $z$ axis, where $v_{\rm rec}$ is the atomic recoil velocity. If the relative velocity is larger than the speed of sound at the condensate center, the collision is supersonic, superfluidity breaks down and a halo of scattered atoms is formed.
The physical properties of this halo are the main subject of our studies.

In our approach we model the process of the halo formation using a time dependent Bogoliubov method, where the field operator is defined as
\begin{equation}\label{Psi-delta}
  \op{\Psi}(\bo{x},t) = \phi(\bo{x},t) + \op{\delta}(\bo{x},t).
\end{equation}
Here, $\phi(\bo{x},t)$ is the condensate wave function governed by the GP equation
\begin{equation}\label{GPeq}
  i\hbar\partial_t\phi(\bo{x},t) = \left(-\frac{\hbar^2}{2m} \nabla^2 + g|\phi(\bo{x},t)|^2\right)\phi(\bo{x},t).
\end{equation}
It is normalized to $N$ -- the number of atoms in the condensate, which remains undepleted during the dynamics. 
The coupling constant $g$ relates to the scattering length $a$ through $g=4\pi\hbar^2a/m$, 
where $m$ is a mass of an atom. 
The field operator  $\op{\delta}(\bo{x},t)$ describes the non-condensed atoms (both the quantum depletion and the scattered halo) and satisfies the following equation
\begin{eqnarray} \label{Bog}
  i\hbar\partial_t\op{\delta}(\bo{x},t)&=&\left(-\frac{\hbar^2}{2m} \nabla^2 + 2g|\phi(\bo{x},t)|^2\right)\op{\delta}(\bo{x},t)\nonumber\\
  &+&g\,\phi^2(\bo{x},t) \op{\delta}^\dagger(\bo{x},t).
\end{eqnarray}
From this equation, we can trace back an effective Bogoliubov Hamiltonian:
\begin{subequations}\label{Heff}\begin{eqnarray}
    \op{H}_{\rm eff} & = & \int d^3\bo{x}\,\dagop{\delta}(\bo{x})\left(-\frac{\hbar^2}{2m} \nabla^2 \right) \op{\delta}(\bo{x}) \label{Heffkin}\\
    &+&2g\int d^3\bo{x}\,|\phi(\bo{x})|^2 \dagop{\delta}(\bo{x})\op{\delta}(\bo{x})\label{Heffmf}\\
    &+&\frac{g}{2} \int d^3\bo{x}\,\phi^2(\bo{x})\,\dagop{\delta}(\bo{x})\dagop{\delta}(\bo{x}) + \text{ h.c. }\label{Heffpair}
\end{eqnarray}\end{subequations}
Line (\ref{Heffkin}) stands for kinetic energy, while line (\ref{Heffmf}) results from the interaction of the
condensate mean-field with the scattered atoms. Finally (\ref{Heffpair}) describes the creation/annihilation process of pairs of non-condensed atoms.

\begin{figure}
\begin{picture}(8,6.0)

	  \put(-0.15,3.0){    \resizebox{4.6cm}{!}{

	      \includegraphics[clip]{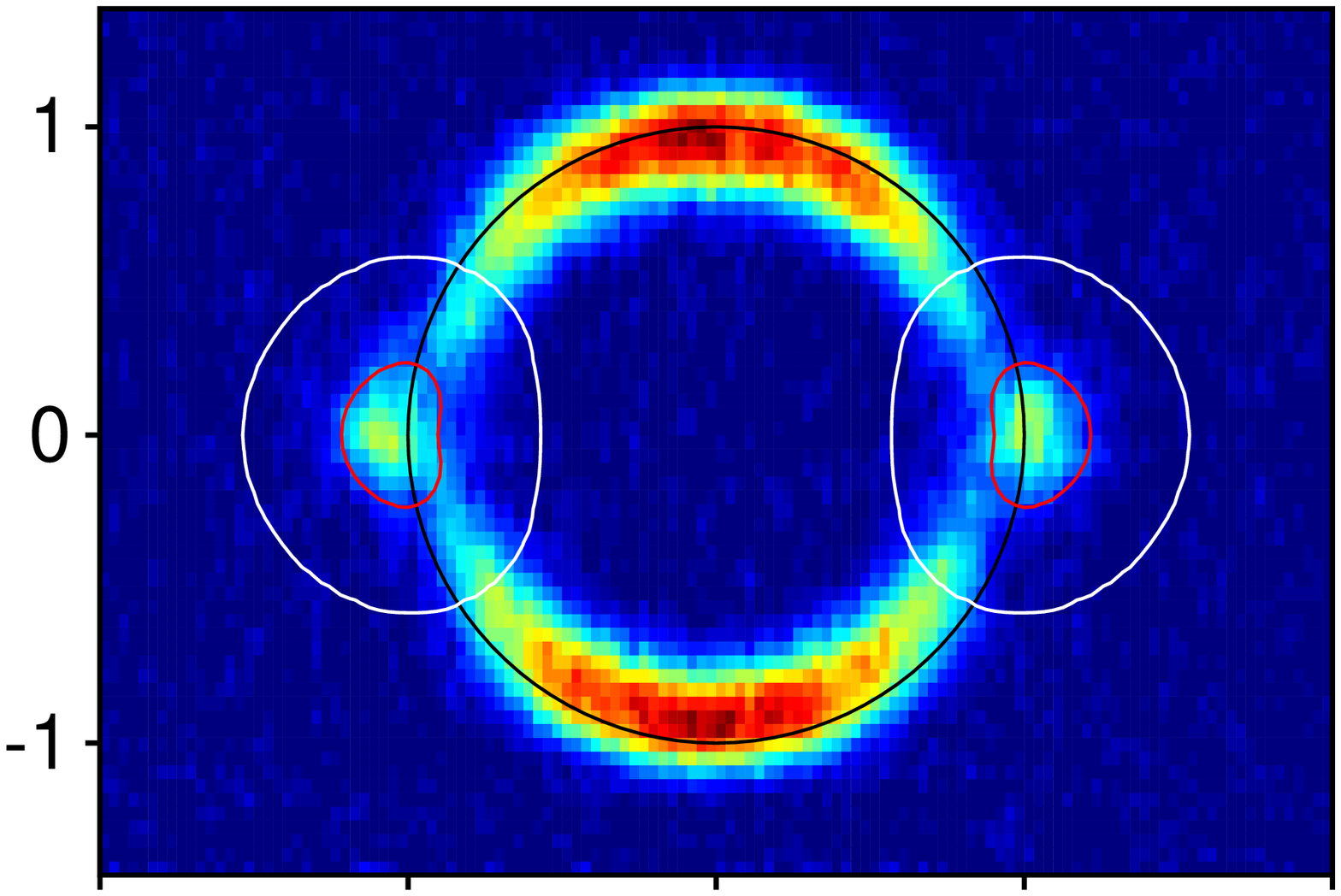}

	  }}

	  \put(0.45,5.6){\Large\color[rgb]{1,1,1}{\textbf{A}}}

	  \put(2.8,5.8){\color[rgb]{1,1,1}{\textbf{Complete}}}

	  \put(3.85,3.0){    \resizebox{4.6cm}{!}{

	      \includegraphics[clip]{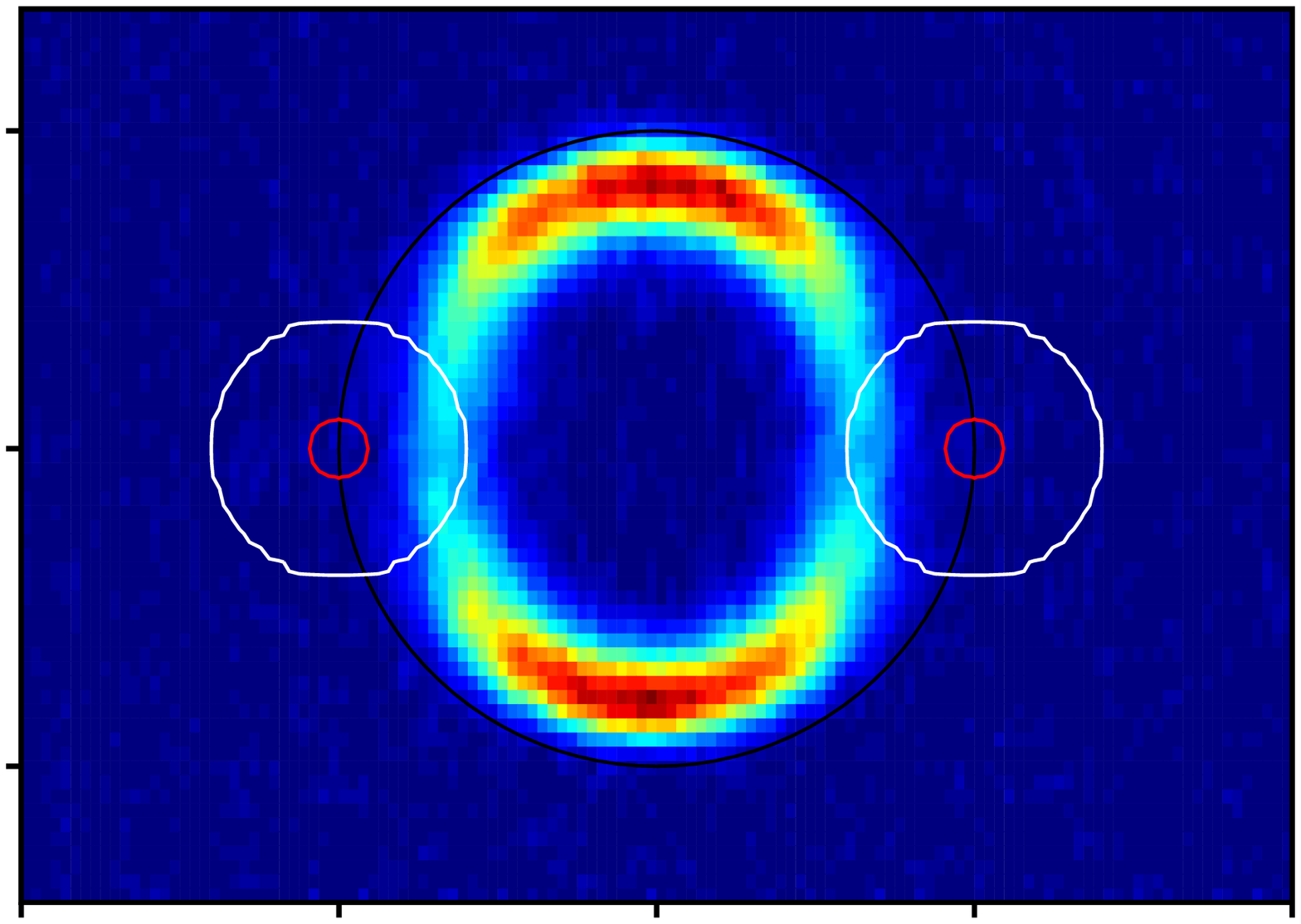}

	  }}

	  \put(4.45,5.6){\Large\color[rgb]{1,1,1}{\textbf{B}}}

	  \put(-0.15,0.1){    \resizebox{4.6cm}{!}{

	      \includegraphics[clip]{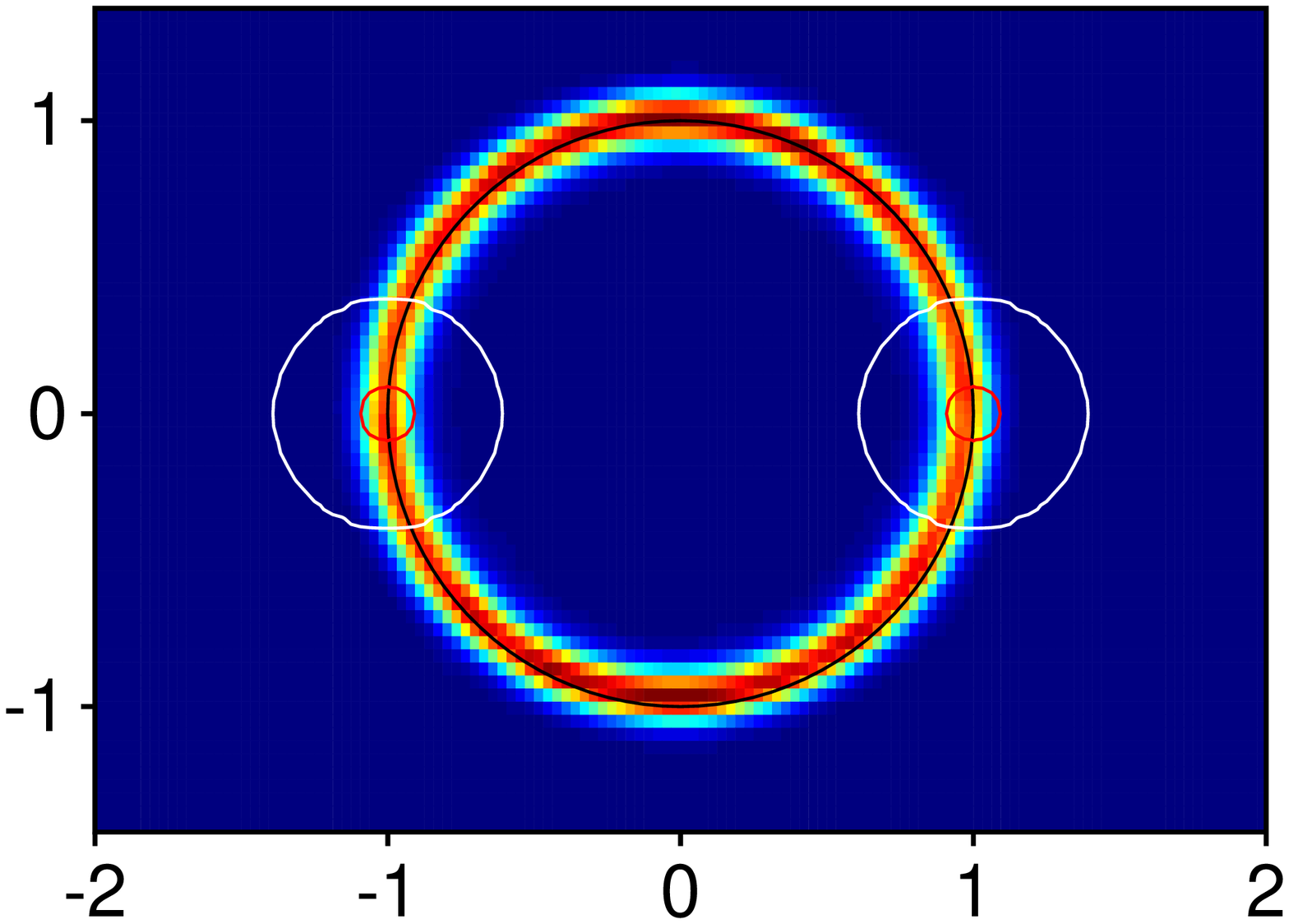}

	  }}

	  \put(0.45,2.7){\Large\color[rgb]{1,1,1}{\textbf{C}}}
	  \put(3.2,2.8){\color[rgb]{1,1,1}{\textbf{RBM}}}

	  \put(3.85,0.1){    \resizebox{4.6cm}{!}{

	      \includegraphics[clip]{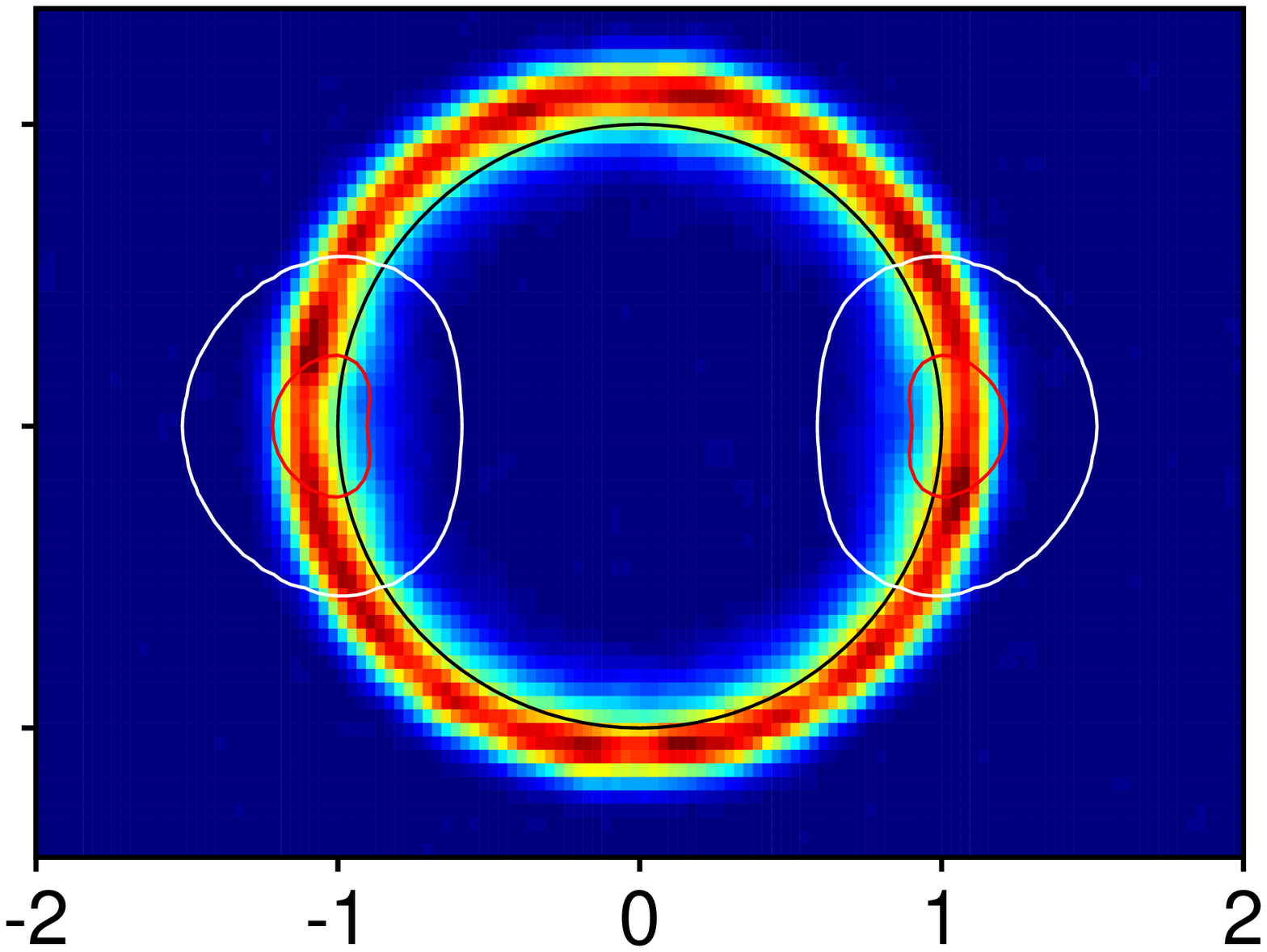}

	  }}

	  \put(4.45,2.7){\Large\color[rgb]{1,1,1}{\textbf{D}}}

	  \put(3.8,-0.05){$k_z/k_0$}

	  \put(-0.2,2.85){\rotatebox{90}{$k_y/k_0$}}

	\end{picture}
	\vspace*{-0.2cm}
      \caption
          {
            Cross-sections through the density of non-condensed atoms in momentum space $n_{\rm nc}(\bo{k})=\langle\dagop{\delta}(\bo{k})\op{\delta}(\bo{k})\rangle$ in the $k_x=0$ plane after the end of the collision, for the bose enhanced case of $\alpha=27.6$, $\beta=31.6$.
            The labels A, B, C, D correspond to the four models introduced in the text.
            Color scale varies between plots.
 	    The black circle shows $|\bo{k}|=k_0$ for reference.
            The red (white) contour is at $0.1$ ($10^{-5}$) of the peak condensate density.
          }\label{big}
\end{figure}


Most salient features of BEC collisions are already found for the case of initially spherically symmetric condensates.
Before the numerical simulation of the collision,
we find the condensate wave-function as a ground state of the Gross-Pitaevskii equation with harmonic trap of frequency $\omega$,
\be\label{GP}	
\mu\phi_0(\bo{x}) = \left(-\frac{\hbar^2}{2m} \nabla^2 + \frac{1}{2}m\omega^2\bo{x}^2 + g|\phi_0(\bo{x})|^2\right)\phi_0(\bo{x}).
\ee
Next, the trap is turned off and at the same time a set of Bragg pulses is applied  \cite{Kozuma99}.
In the center-of-mass frame, the initial state is then
\be\label{GPic}	
\phi(\bo{x},0) = A\phi_0(\bo{x}) \left(e^{ik_0 z}+e^{-ik_0 z}\right),
\ee
where $A$ is the normalization constant and $k_0=mv_{\rm rec}/\hbar$ is the wave-vector associated with the recoil velocity.
The initial state of the non-condensed part is taken to be a vacuum
\begin{equation}\label{Initop}
  \op{\delta}(\bo{x},0) |0\rangle = 0.
\end{equation}
From this initial state the system evolves according to equations (\ref{GPeq}) and (\ref{Bog}).

\subsection{Dimensionless parameters}
\label{PARAMS}

We point out that in our approximate description (Bogoliubov method), there is a universal scaling such that the non-condensed field $\op{\delta}(\bo{x})$ is identical in all systems having the same value of $gN$ (or $aN$, where $a$ is a scattering length). Hence the dynamics is described by the length scale $aN$ rather than by the scattering length $a$ or the number of particles $N$ separately. Other important length scales are $a_{ho}=\sqrt{\hbar/m\omega}$ -- the harmonic oscillator length,  and $1/k_0$. Alternatively, since there is no trap for $t>0$, we can use the width of the initial condensate $\sigma$ instead of $a_{ho}$. If we apply the Thomas-Fermi approximation  \cite{Dalfovo99}, a good choice of characteristic width $\sigma$ is the Thomas Fermi radius $R_{TF} = (15 N g/ 4\pi m \omega^2)^{1/5}$.
In conclusion: there are three relevant length scales, hence our system can be characterized by two dimensionless parameters.
We choose: $\beta=k_0\sigma$ and $\alpha=aN/\sigma$ (see \cite{Zin05,Zin06,Chwedenczuk06,Chwedenczuk08}).

The first parameter $\beta$ is independent of interactions and has a kinetic character. It can be viewed as the number of fringes created by the Bragg pulses on the initial condensate,
or as a ratio of the dispersion time scale, $m \sigma^2/\hbar$ to the collision timescale, $(m \sigma)/(\hbar k_0)$, see for instance \cite{Zin05,Zin06,Chwedenczuk06,Chwedenczuk08}.
In realistic situations $\beta\gg1$.

The second parameter $\alpha$ is proportional to the ratio of the interaction energy per particle $g N/\sigma^3$ to the kinetic energy per particle in
the initial condensate $\hbar^2/(2m \sigma^2)$. It is related to $\alpha^{\rm(past)}=mgN/(\sigma\hbar^2\pi^{3/2})$ which has been used previously \cite{Zin05,Zin06,Chwedenczuk06,Chwedenczuk08}, but differs by a numerical factor: $\alpha =  (\sqrt{\pi}/4)\alpha^{\rm(past)}$.
The ratio between $\alpha$ and $\beta$ quantifies the strength and nature of scattering. 
It has been shown that Bose enhancement of scattering occurs for $\alpha \gtrsim \beta$  \cite{Zin06}.

\subsection{Simulation of dynamics using the STAB method}
\label{METHOD}

The Bogoliubov equation (\ref{Bog}) can be solved numerically using the positive-P representation method  \cite{stab}, where the field operator $\op{\delta}$ is replaced with two complex-number fields
$\psi(\bo{x})$ and $\wt{\psi}(\bo{x})$. The dynamics of the system is governed by a pair of stochastic Ito equations,
\begin{widetext}
  \begin{subequations}\label{STABeq}\begin{eqnarray}
      i\hbar\partial_t\psi({\bo{x}},t) &=& \left(-\frac{\hbar^2}{2m} \nabla^2 + 2g|\phi({\bo{x}},t)|^2\right)\psi({\bo{x}},t) + g\phi({\bo{x}},t)^2\wt{\psi}({\bo{x}},t)^* +
      \sqrt{i\hbar g}\,\phi({\bo{x}},t) \xi({\bo{x}},t),\label{STABpsi}\\
      i\hbar\partial_t\wt{\psi}({\bo{x}},t)&=& \left(-\frac{\hbar^2}{2m} \nabla^2 + 2g|\phi({\bo{x}},t)|^2\right)\wt{\psi}({\bo{x}},t) + g\phi({\bo{x}},t)^2\psi({\bo{x}},t)^* +
      \sqrt{i\hbar g}\,\phi({\bo{x}},t) \wt{\xi}({\bo{x}},t).\label{STABpsit}
\end{eqnarray}\end{subequations}\end{widetext}
Here $\xi({\bo{x}},t)$ and $\wt{\xi}({\bo{x}},t)$ are delta-correlated, independent, real stochastic noise fields with zero mean. The second moments are equal to
\begin{eqnarray*}
  &&\langle \xi({\bo{x}},t)\wt{\xi}({\bo{x}}',t') \rangle =0\ \ \ \ \ \ \mathrm{and}\ \ \ \ \ \ \\
  &&\langle \xi({\bo{x}},t)\xi({\bo{x}}',t') \rangle
  = \langle \wt{\xi}({\bo{x}},t)\wt{\xi}({\bo{x}}',t') \rangle
  =\delta({\bo{x}}-{\bo{x}}')\delta(t-t').
  \end{eqnarray*}
Numerically, $\xi$ and $\wt{\xi}$ are approximated by real gaussian random variables of variance $1/(\Delta t\Delta V)$ that are independent at each point
at the computational lattice (of volume $\Delta V$), and at each time step of length $\Delta t$.

\begin{figure}
	\begin{picture}(8,6.0)

	  \put(-0.15,3.0){    \resizebox{4.6cm}{!}{

	      \includegraphics[clip]{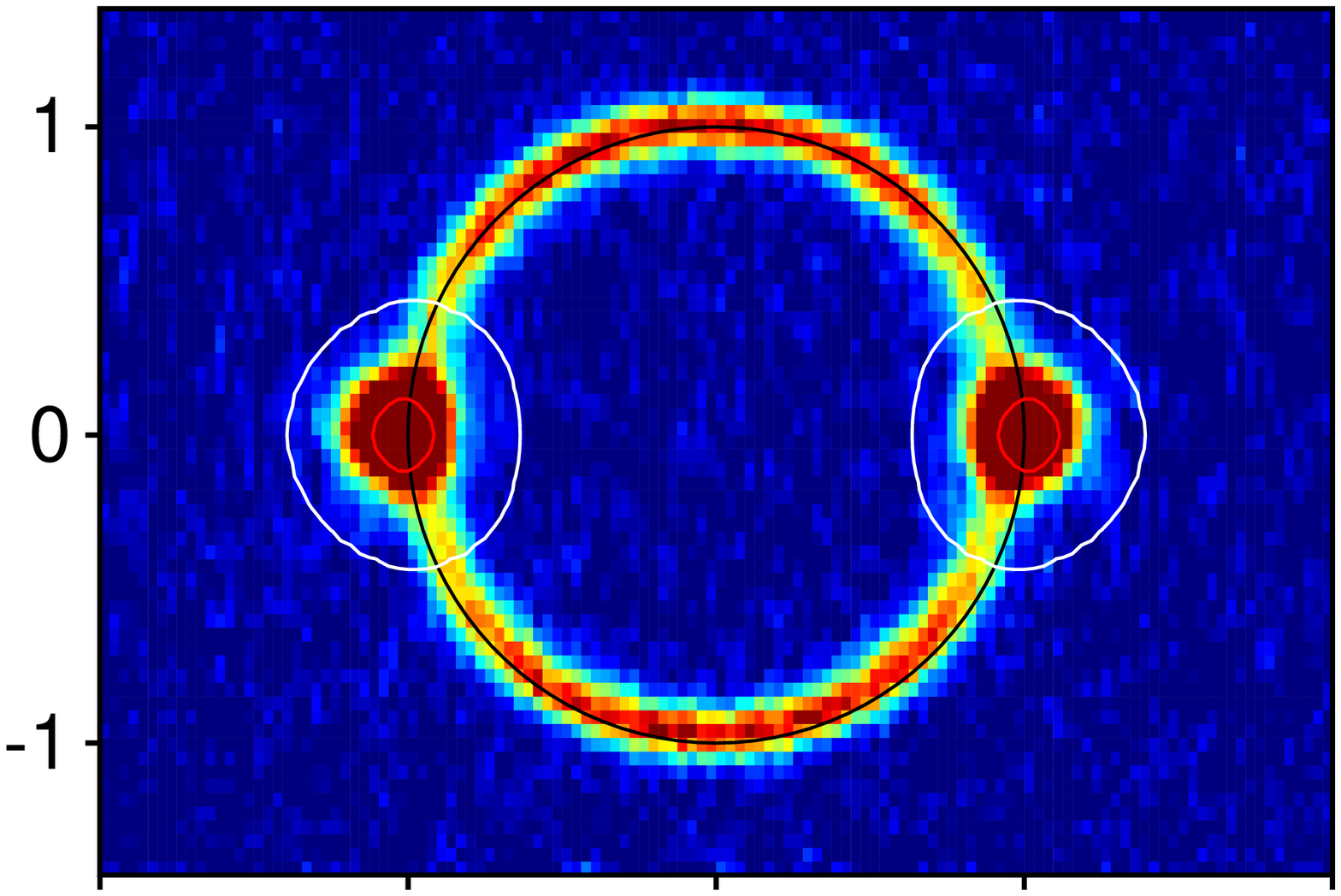}

	  }}

	  \put(0.45,5.6){\Large\color[rgb]{1,1,1}{\textbf{A}}}

	  \put(2.8,5.8){\color[rgb]{1,1,1}{\textbf{Complete}}}

	  \put(3.85,3.0){    \resizebox{4.6cm}{!}{

	      \includegraphics[clip]{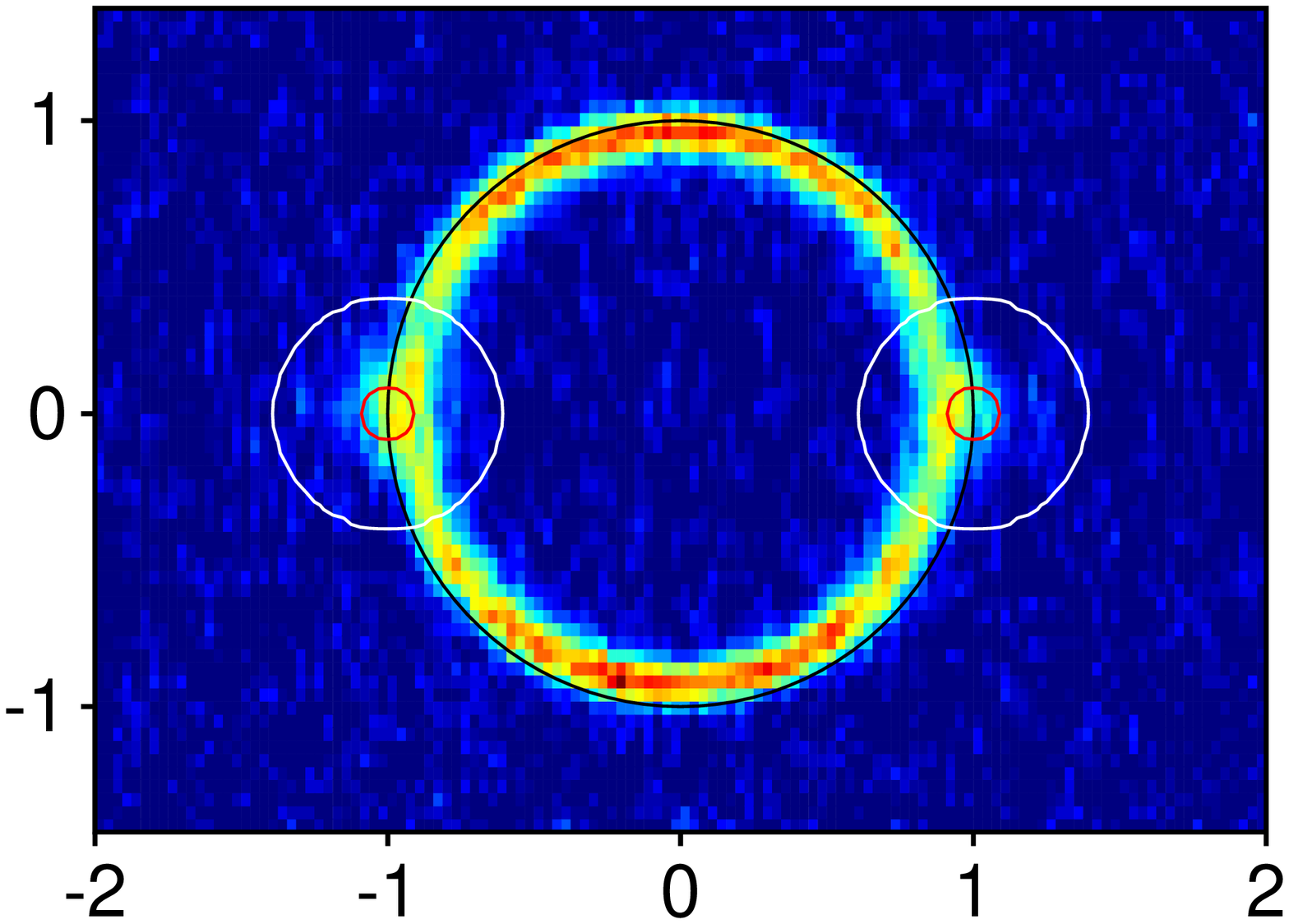}

	  }}

	  \put(4.45,5.6){\Large\color[rgb]{1,1,1}{\textbf{B}}}

	  \put(-0.15,0.1){    \resizebox{4.6cm}{!}{

	      \includegraphics[clip]{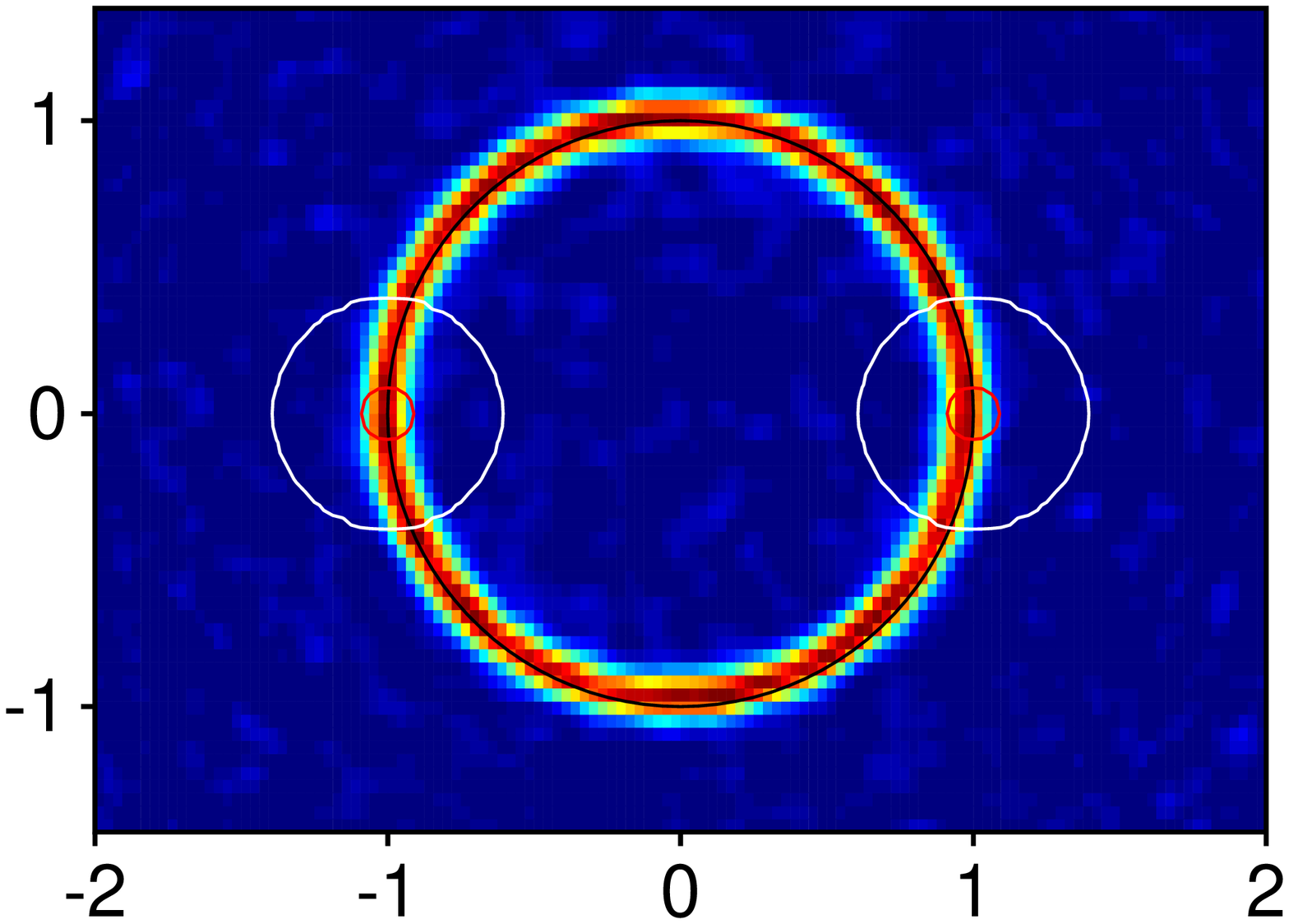}

	  }}

	  \put(0.45,2.7){\Large\color[rgb]{1,1,1}{\textbf{C}}}

	  \put(3.2,2.8){\color[rgb]{1,1,1}{\textbf{RBM}}}

	  \put(3.85,0.1){    \resizebox{4.6cm}{!}{

	      \includegraphics[clip]{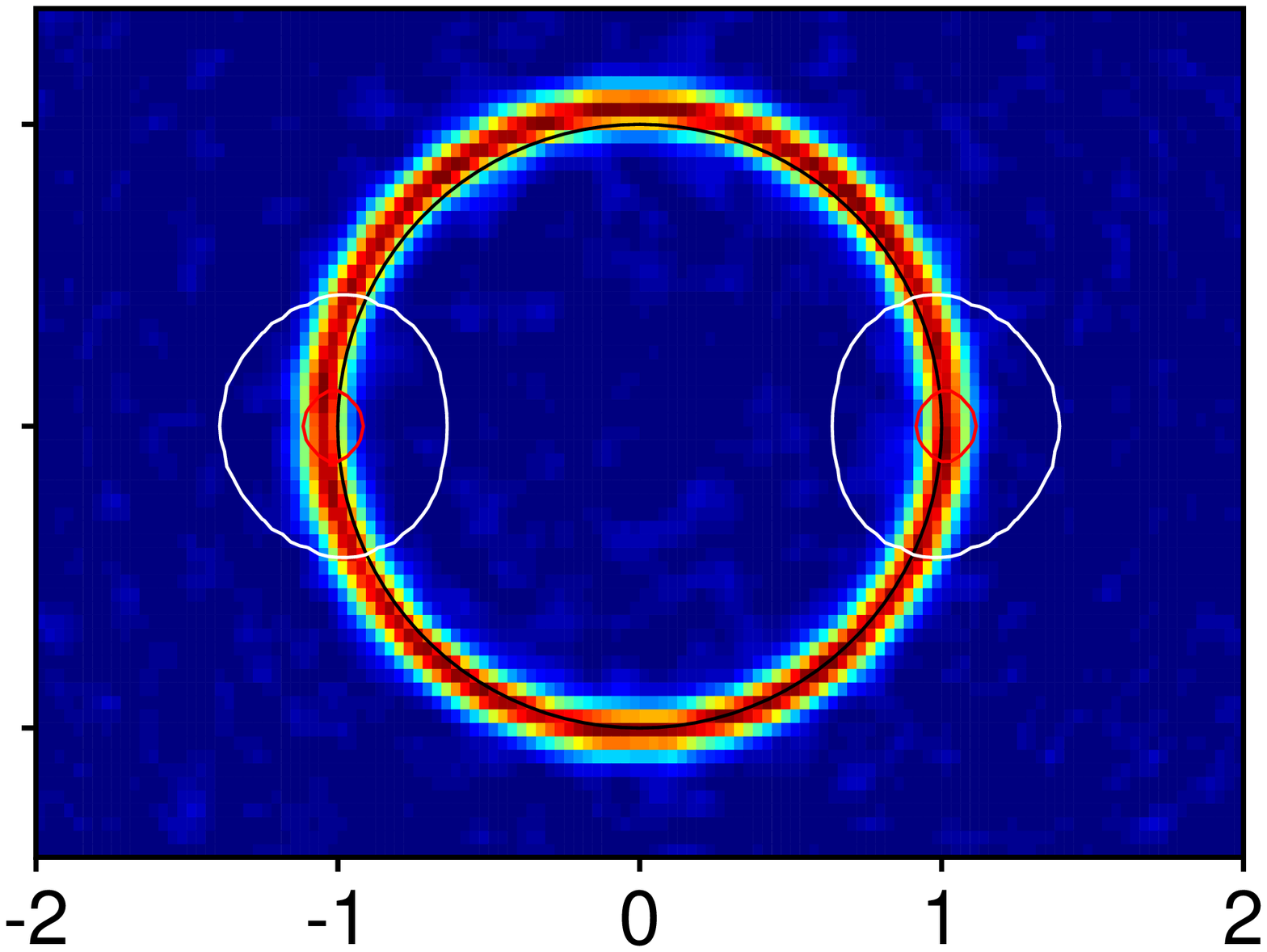}

	  }}

	  \put(4.45,2.7){\Large\color[rgb]{1,1,1}{\textbf{D}}}

	  \put(3.8,-0.05){$k_z/k_0$}

	  \put(-0.2,2.85){\rotatebox{90}{$k_y/k_0$}}

	\end{picture}

	\vspace*{-0.2cm}
      \caption
          {
		As Figure~\ref{big}, but for the case of $\alpha=9.2$, $\beta=31.6$ where bosonic stimulation of scattering is negligible.
          }\label{small}
\end{figure}

Any physical quantity is obtained by substituting $\dagop{\delta}\to\wt{\psi}^*$ and $\op{\delta}\to\psi$ and
changing from a quantum average of the normally-ordered operator to a stochastic average  \cite{Deuar02},
\begin{equation*}\label{norm-order}
  \Big\langle\prod_j\dagop{\delta}(\bo{x}_j)\prod_k\op{\delta}({\bo{x}}_k)\Big\rangle
  = \lim_{m_r\to\infty}\Big\langle\prod_{j,k} \wt{\psi}({\bo{x}}_j)^*\psi({\bo{x}}_k)\Big\rangle_{st}.
\end{equation*}
The braces $\langle\cdot\rangle_{st}$ denote the statistical average over $m_r$ realizations.
Observables in k-space follow from the Fourier transformation. For example, the one-particle density reads
\begin{equation*}
  \rho_1(\bo{k},\bo{k}') = \phi(\bo{k})^*\phi(\bo{k}') + \textrm{Re}\langle\wt{\psi}(\bo{k})^*\psi(\bo{k}')\rangle_{st}.\label{obs-rho1-primary}
\end{equation*}

The stochastic equations (\ref{STABeq})  strictly reproduce the full quantum dynamics described by $\op{H}_{\rm eff}$ when the number of samples tends to infinity.
For finite sample sizes, one obtains an estimator for the full quantum dynamics, with an uncertainty that is calculated by standard methods \cite{PHDeuar}.

\subsection{Partially reduced Bogoliubov models}

A fortuitous ``side-effect'' of the method formulated above is the possibility to systematically add or remove parts of the effective Hamiltonian (\ref{Heff}) and GP equation (\ref{GPeq}) to understand
their impact on the dynamics of the $\hat\delta$ field. This includes a numerical simulation of the RBM.To do this, we first write the condensate wavefunction as a sum of left- and
right-moving wavepackets,
\begin{equation}\label{phiLR}
  \phi(\bo{x},t)=\phi_L(\bo{x},t) + \phi_R(\bo{x},t),
\end{equation}
where $\phi_{L/R}(\bo{x},0)=A\phi_0(\bo{x})e^{\pm ik_0 z}$ with the GP equation
\begin{subequations}\label{GPeqLR}
  \begin{eqnarray}
    i\hbar\partial_t\phi_{L/R}(\bo{x},t)&=& \left(-\frac{\hbar^2}{2m} \nabla^2 \label{GPeqLRkin}\right.\\
    && + g\,|\phi_{L/R}(\bo{x},t)|^2\label{GPeqLRbroad}\\
    && + 2g\,|\phi_{R/L}(\bo{x},t)|^2\label{GPeqLRpush}\\
    && + g\,\phi_{R/L}(\bo{x},t)^*\phi_{L/R}(\bo{x},t)\label{GPeqLRintf}\\
    &&\left.\hspace*{14mm}\phantom{\frac{\hbar^2}{2m}}\right)\phi_{L/R}(\bo{x},t).\nonu
  \end{eqnarray}
\end{subequations}
The terms proportional to the coupling constant $g$ can be interpreted as the self-interaction of the wavepacket (\ref{GPeqLRbroad})
and cross-interaction between different wavepackets (\ref{GPeqLRpush}) respectively. The remaining processes are in line (\ref{GPeqLRintf}).

Next, the decomposition (\ref{phiLR}) is put into the Bogoliubov Hamiltonian (\ref{Heff}) giving
\begin{subequations}\label{HeffLR}
  \begin{eqnarray}
    \op{H}_{\rm eff} & = &  \int d^3\bo{x}\,\dagop{\delta}(\bo{x}) \left( -\frac{\hbar^2}{2m} \nabla^2 \right) \op{\delta}(\bo{x}) \label{HeffLRkin}\\
    &&+ 2g \int d^3\bo{x}\, \left|\phi_L(\bo{x})+\phi_R(\bo{x})\right|^2 \dagop{\delta}(\bo{x})\op{\delta}(\bo{x})\label{HeffLRmf}\\
    &&+ g \int d^3\bo{x}\, \phi_L(\bo{x})\phi_R(\bo{x}) \dagop{\delta}(\bo{x})\dagop{\delta}(\bo{x}) + \text{ h.c. }\label{HeffLRpair}\\
    &&\hspace*{-10mm}+ \frac{g}{2} \int d^3\bo{x}\, \left(\phi_L(\bo{x})^2+\phi_R(\bo{x})^2\right) \dagop{\delta}(\bo{x})\dagop{\delta}(\bo{x}) + \text{ h.c. }\qquad\label{HeffLRoffres}
  \end{eqnarray}
\end{subequations}
The resonant term (\ref{HeffLRpair}) governs the process where two atoms, one from $\phi_R(\bo{x})$ and one from $\phi_L(\bo{x})$, 
collide and elastically scatter into the halo localized
around the radius $|\bo{k}|\approx k_0$ as in the usual RBM. The line (\ref{HeffLRoffres}) leads to creation of two atoms in the $\hat\delta$ field originating from one
wave-packet, and we call this here ``on-condensate pairing''. This incudes quantum depletion processes. 

Note that the specific form of Equations
(\ref{GPeqLR}--\ref{HeffLR}) allows for removal of chosen terms. This way, one can inspect their role in the dynamics of the system.
\begin{figure}
  \includegraphics[clip,scale=0.32]{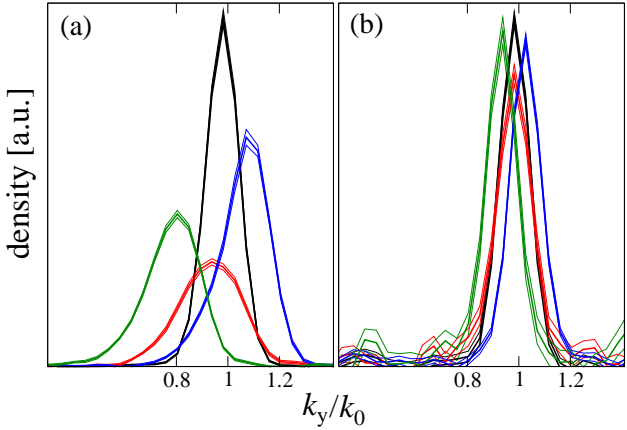}
  \caption
      {
        \label{Fig-halo}
	Cross-sections through the halo of scattered atoms along the $k_y$ axis after the end of the collision for $\alpha=27.6$ \textbf{(a)} and
        $\alpha=9.2$ \textbf{(b)}.
	The curves are calculated using the models A (red), B (green), C (black) and D (blue).
        Triple lines indicate the best statistical estimate and the $\pm1\sigma$ statistical uncertainties in the mean.
      }
\end{figure}
For instance, we can neglect all terms but the kinetic energy (\ref{HeffLRkin}) and pair production (\ref{HeffLRpair}) in the Bogoliubov Hamiltonian. This simplification we call the Pair Production (PP) dynamics. Within this approximation, the Positive P formulation takes the form:
\begin{subequations}
  \label{reducedBogH}
  \begin{eqnarray}
    i\hbar\partial_t\psi(\bo{x},t) &=&  -\frac{\hbar^2 \nabla^2}{2m}\psi(\bo{x},t) + 2g\phi_L(\bo{x},t)\phi_R(\bo{x},t)\wt{\psi}(\bo{x},t)^*\nonumber\\
    &+&\sqrt{2i\hbar g\phi_L(\bo{x},t)\phi_R(\bo{x},t)}\,\xi(\bo{x},t),\label{reducedBogPsi}\\
    i\hbar\partial_t\wt{\psi}(\bo{x},t) &=&  -\frac{\hbar^2 \nabla^2}{2m}\wt{\psi}(\bo{x},t) + 2g\phi_L(\bo{x},t)\phi_R(\bo{x},t)\psi(\bo{x},t)^*\nonumber\\
    &+&\sqrt{2i\hbar g\phi_L(\bo{x},t)\phi_R(\bo{x},t)}\,\wt{\xi}(\bo{x},t).\label{reducedBogPsit}
  \end{eqnarray}
\end{subequations}
We can also simplify the GP equation (\ref{HeffLR}) neglecting all the nonlinear terms and approximating the kinetic energy operator. As a result of this treatment we obtain
the stiff movement (SM) of the counter-propagating wave-packets,
\begin{subequations}
  \label{reducedBogGP}
  \begin{eqnarray}
    i\hbar\partial_t\phi_L(\bo{x},t)&=& -\frac{\hbar^2k_0}{2m}\left(k_0-2i\frac{\partial}{\partial z}\right) \phi_L(\bo{x},t)\qquad\label{reducedBoggpL}\\
    i\hbar\partial_t\phi_R(\bo{x},t)&=& -\frac{\hbar^2k_0}{2m}\left(k_0+2i\frac{\partial}{\partial z}\right) \phi_R(\bo{x},t).\qquad\label{reducedBoggpR}
  \end{eqnarray}
\end{subequations}

The RBM, described in literature, consists of both the PP and SM simplifications. Also, using alternative
combinations of the approximations described above, we can generate four different models:\\
\begin{description}
\item[A] GP equation (\ref{GPeq}) + full Bogoliubov (\ref{STABeq})
\item[B] SM  (\ref{reducedBogGP}) recombined to (\ref{phiLR}) + full Bogoliubov (\ref{STABeq}).
\item[C] SM (\ref{reducedBogGP}) + PP  (\ref{reducedBogH}) equation $\rightarrow$ we call it RBM
\item[D] GP equation (reduced as in (\ref{GPeqLR})) + PP (\ref{reducedBogH}) equation.
\end{description}

In the following section we compare the distributions of scattered atoms obtained using the four above models.

\section{Results}
\subsection{The shape of the halo}
\label{EG1}

We first investigate the dynamics of the system in two characteristic cases, varying only $\alpha$, while $\beta=31.6$ is kept constant.
It has been previously shown, that for Gaussian colliding clouds, significant Bose enhancement of scattering occurs when $\alpha^{\rm(past)}/\beta\approx2$ or greater, i.e.
$\alpha\gtrsim\beta$. Figures \ref{big} and \ref{small}
show the cross-section through the halo of scattered atoms in the $k_x=0$ plane at the end of the collision, for the case of appreciable ($\alpha=27.6$) and negligible ($\alpha=9.2$) bosonic enhancement of scattering.
Density profiles of the halo in these two cases are shown in Fig.~\ref{Fig-halo}.

When the dynamics of the $\hat\delta$ field is described by the simplified models C and D, the density of scattered atoms is spherically
symmetric. On the other hand, this symmetry is lost within models A and B, where the halo of atoms is weakened near the condensates.
This effect can be attributed primarily to the mean field (\ref{HeffLRmf}), with some enhancement resulting from (\ref{HeffLRoffres}).
Notice that for larger interaction strength -- and consequently larger $\alpha$ -- these phenomena are more pronounced.
Moreover, models A and B predict some non-condensed atoms appearing on top of the BECs (in the same location as the BECs).
The latter effect results from the term (\ref{HeffLRoffres}) only, and it as related to quantum depletion.

\subsection{Position of the halo center}
\label{EG}

In the momentum space we denote location of the maximum of the halo density by $k_{\rm max}$. This parameter, briefly discussed in \cite{Krachmalnicoff10},
varies between models A-D, as can be seen in Figs.~\ref{big} and \ref{small} and Fig.~\ref{Fig-halo} in more detail.
The shift of the position of the halo can be explained when the BECs are modelled with two counter-propagating plane-waves \cite{Krachmalnicoff10,ZinUnpubl}, 
and using energy conservation argument. 
The energy of a particle released from the condensates depends on the form of the GP equation we use to describe it.
The full GP equation, with $n$ - the mean density of the system, gives $\frac{\hbar^2k_0^2}{2m} + \frac{3}{2}gn $, while the SM (\ref{reducedBogGP}) gives a purely kinetic value 
$\frac{\hbar^2k_0^2}{2m}$.
On the other hand the energy needed to place a particle in a noncondensed mode with momentum $\hbar k$ is equal to 
$\frac{\hbar^2k^2}{2m}+2gn$ in the case of full Bogoliubov and $\frac{\hbar^2k^2}{2m}$ in the case of the PP equation (\ref{reducedBogH}). Hence, using energy conservation we obtain
\begin{description}
\item[A] \qquad$\frac{\hbar^2k^2}{2m} = \frac{\hbar^2k_0^2}{2m} - \frac{1}{2}gn$
\item[B] \qquad$\frac{\hbar^2k^2}{2m}=\frac{\hbar^2k_0^2}{2m} - 2gn$
\item[C] \qquad$\frac{\hbar^2k^2}{2m}=\frac{\hbar^2k_0^2}{2m}$
\item[D] \qquad$\frac{\hbar^2k^2}{2m}=\frac{\hbar^2k_0^2}{2m} + \frac{3}{2}gn$ 
\end{description}
This simplified model gives quantitative agreement with the results presented in Figure \ref{Fig-halo}.

\begin{figure}
  \includegraphics[clip,scale=0.32]{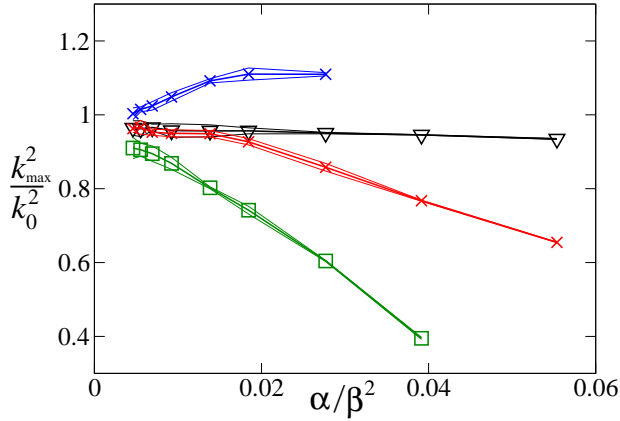}
  \caption
      {
        \label{Fig-pos}
	Dependence of the halo peak position $k_{\rm max}$ on $\alpha/\beta^2$, calculated with
        A (red circles), B (green squares), C (black triangles) and D (blue $\times$).
        Error bars indicate the  $\pm1\sigma$ statistical uncertainties in the mean.
      }
\end{figure}

Notice that in all models $(k/k_0)^2$ is a linear function of the parameter $gn\,\frac{2m}{\hbar^2k_0^2}$. In order to generalize this dependence to the case of non-uniform condensates,
we replace $n$ with half of the maximal density, which in the Thomas-Fermi approximation is $n_{max}=(15/8\pi) N /R_{TF}^3$. Recalling the definitions
of $\alpha$ and $\beta$, we obtain the shift proportional to $\alpha/\beta^2$. To verify this conjecture, 
in Fig.\ref{Fig-pos} we plot $(k_{\mathrm{max}}/k_0)^2$ as a function of $\alpha/\beta^2$ for all four models. 
We observe that for growing $\alpha/\beta^2$ (and thus growing $\alpha$, as we keep constant
$\beta=31.6$), there is some deviation from the linear behaviour for models A and D. What these two models have in common is the full evolution of the condensates,
which might make the picture of plane wave collisions with a time-independent density difficult to uphold. 

Finally, we note that for small enough $\alpha$, the discrepancy between the RBM and full Bogoliubov treatment is negligible, as expected.
The region of agreement ($\alpha\lesssim\beta$) matches approximately the region where stimulated scattering (Bose enhancement) is absent, 
although investigation with various values of $\beta$ would be necessary to confirm or refute a link.

\section{Conclusions}
\label{CONCL}
We have demonstrated how different approximations of the full Bogoliubov equation dramatically influence the results of numerical simulations of BEC collisions. 
Although the simpliest RBM approach predicts a spherical density of scattered atoms, more involved models show significant discrepancies from this distribution. These differences can be understood as a result of the action of
mean-field terms in the GP and Bogoliubov equations. Also, the position of the density maximum changes from model to model. This effect, which was investigated  when
the BECs are approximated by the plane-waves, is here interpreted for non-uniform condensates. 
The RBM model is quantitatively correct when the interaction strength, as quantified by $\alpha$, is sufficiently small.

\begin{acknowledgments}
We acknowledge fruitful discussions with Chris Westbrook, Denis Boiron and Karen Kheruntsyan. M. T. acknowledges support of Polish Government Research Grants for 2007-2011,
P. D. support by the EU contract PERG06-GA-2009-256291 and J. C. was supported by Foundation for Polish Science International TEAM Programme co-financed by the EU European Regional Development Fund.
\end{acknowledgments}

\newcommand{\PRL}[1]{Phys. Rev. Lett.~\textbf{#1}}
\newcommand{\PRA}[1]{Phys. Rev.~A~\textbf{#1}}

\end{document}